\documentclass[%
 reprint,
superscriptaddress,
 amsmath,amssymb,
 aps,
]{revtex4-2}

\usepackage[english]{babel}

\usepackage{amsmath}
\usepackage{amsfonts}
\usepackage{graphicx}
\usepackage[colorlinks=true, allcolors=blue]{hyperref}

\newcommand{\ie}{{\it i.e.~}}

\begin{document}

\title{On the efficiency of an autonomous dynamic Szilard engine \\ operating on a single active particle}

\author{Luca Cocconi}
\email{luca.cocconi@ds.mpg.de}
\affiliation{Max Planck Institute for Dynamics and Self-Organization, G{\"o}ttingen 37073, Germany}

\author{Letian Chen}
\affiliation{Department of Mathematics and Centre of Complexity Science, Imperial College London, South Kensington, London SW7 2BZ, United Kingdom}

\date{\today}


\begin{abstract}
The Szilard engine stands as a compelling illustration of the intricate interplay between information and thermodynamics. While at thermodynamic equilibrium, the apparent breach of the second law of thermodynamics was reconciled by Landauer and Bennett's insights into memory writing and erasure, recent extensions of these concepts into regimes featuring active fluctuations have unveiled the prospect of exceeding Landauer's bound, capitalizing on information to divert free energy from dissipation towards useful work.
To explore this question further, we investigate an autonomous dynamic Szilard engine, addressing the thermodynamic consistency of work extraction and measurement costs by extending the phase space to incorporate an auxiliary system, which plays the role of an explicit measurement device. The nonreciprocal coupling between active particle and measurement device introduces a feedback control loop, and the cost of measurement is quantified through excess entropy production. The study considers different measurement scenarios, highlighting the role of measurement precision in determining engine efficiency. 
\end{abstract}

\maketitle

\section{Introduction}

Starting with Maxwell's seminal \emph{Gedankenexperiment}, in which the existence of an entity capable of sorting individual molecules according to their velocity would seemingly lead to a violation of the second law of thermodynamics \cite{knott1911quote}, the physical nature of information and its interplay with thermodynamics has undergone much scrutiny. 
This interplay finds its most vivid illustration in another thought experiment, this time concocted by Szilard \cite{szilard1964decrease}. In Szilard's \emph{engine}, a partition is inserted at the mid-plane of a box containing a single gas particle upon a binary measurement of the position of the particle relative to the midplane. The volume of the empty half of the box is then reduced at no energetic cost, resulting, once the partition is removed, in an increase in free energy of the one-particle gas which is subsequently converted into up to $k_B T \ln 2$ Joules of useful work via isothermal expansion. Later work by Landauer and Bennet \cite{landauer1961irreversibility,bennett2003notes} offered a resolution to the seeming paradox by noting that, particularly when operated cyclically, all such information engines rely on the writing onto, storing in and eventual erasure from physical memory, thus demanding irreversible manipulations which are bound to generate entropy {as a by-product}. In short, thermodynamic consistency is recovered upon expanding the phase space to include the \emph{daemon} itself \cite{strasberg2013thermodynamics,shiraishi2015role,mandal2012work,horowitz2013imitating,Vedral2009RevModPhys,Sagawa2015Natphy}. Theoretical efforts to clarify where precisely dissipation occurs continue to this day \cite{daimer2023physical,ouldridge2018power,leff2002maxwell,Jarzynski2013refrigerator}, in part stimulated by experimental work \cite{jun2014high,sahainfo2023,saha2022bayesian,saha2021maximizing}. 

The traditional Szilard engine, in its various implementations, is assumed to operate on systems coupled to an equilibrium thermal reservoir and it is indeed the measurement-mediated rectification of thermal fluctuations induced by this coupling that renders work extraction possible. It is thus natural to wonder how such an engine would perform when allowed to operate on out-of-equilibrium processes, e.g.\ active particles, which are subject to non-negligible active fluctuations with {macroscopic} persistence times. Remarkably, as demonstrated in recent theoretical \cite{malgaretti2022} and experimental \cite{sahainfo2023} works, such activity allows for the violation of Landauer's bound, potentially granting access to efficiencies far exceeding the equilibrium limit.  Qualitatively, this can be understood as a consequence of information being used in this case to redirect part of the free energy that would otherwise be dissipated as heat (\ie entropy production) into useful work, rather than to extract the latter directly from the heat bath. 

In particular, the authors of Ref.~\cite{malgaretti2022} {introduced} a hypothetical \emph{dynamic} Szilard engine which, rather than operating in a quasi-static regime, exploits the finite correlation time of the velocity of an active Brownian particle (ABP). Their protocol consists of repeatedly measuring the particle's position and direction of motion, subsequently placing a piston that will exert a force opposite to the active self-propulsion, eventually resulting in positive work against the former. The cost of measurement is estimated as $\mathcal{M} \simeq -k_B T \ln(\delta/2)$ by analogy with the thermal case, with $\delta$ the precision {error} of the position measurement normalized by the system size. {A unified stochastic thermodynamic treatment encompassing the quantification of both work extraction and measurement costs in such active information engines in the spirit of Refs.~\cite{strasberg2013thermodynamics,shiraishi2015role,mandal2012work,horowitz2013imitating,Vedral2009RevModPhys,Sagawa2015Natphy}, however, is yet to be established.}\\

\begin{figure}
    \centering
    \includegraphics[width=\columnwidth]{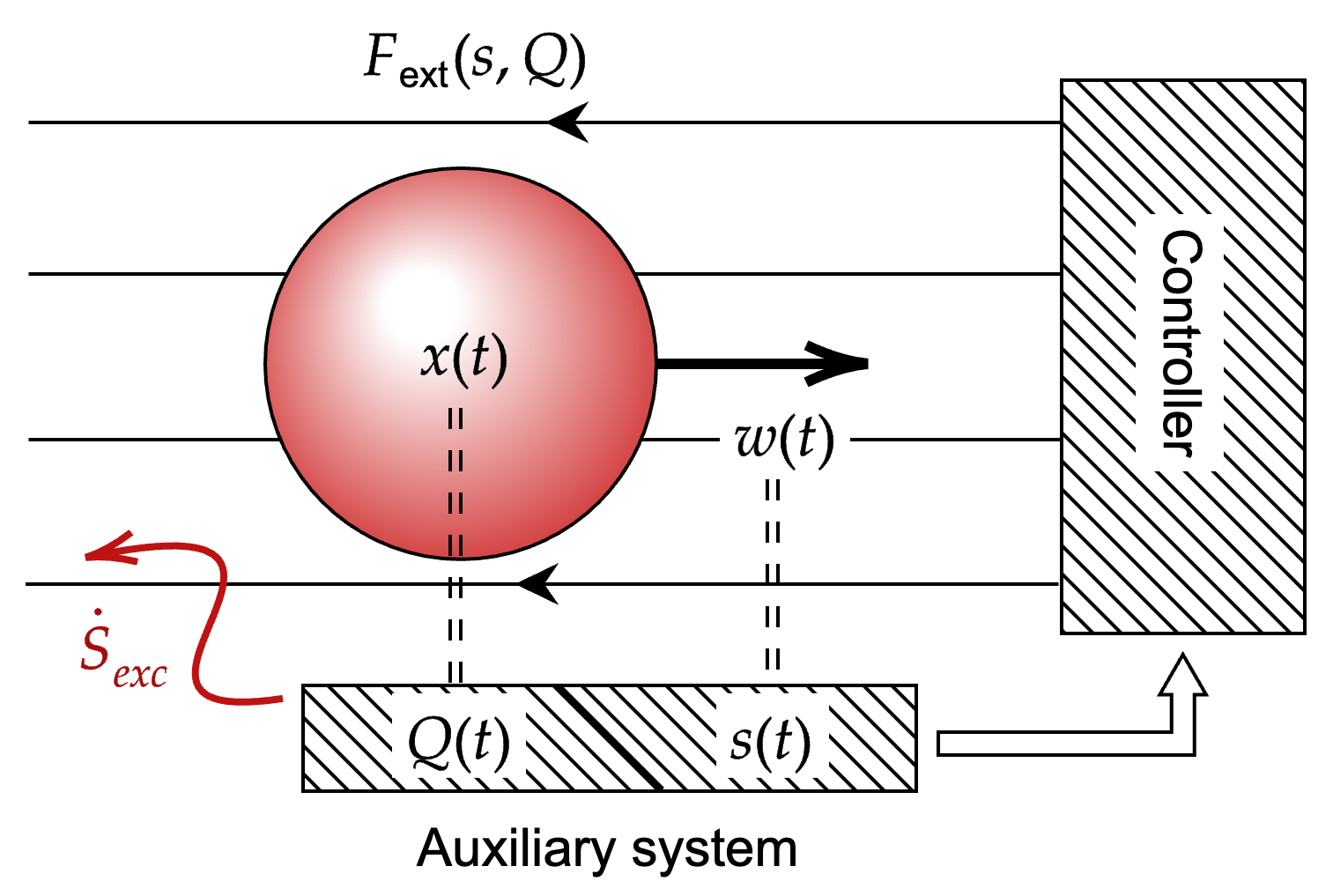}
    \caption{Schematic representation of the dynamic Szilard engine considered in this work. A single active particle undergoing run-and-tumble motion is subject to a time-dependent force $F_{\rm ext}$ applied by an external controller. The force is modulated based on the current state of auxiliary processes $s$ or $Q$, which are coupled to the internal self-propulsion state $w$ and particle position $x$, respectively. The nonequilibrium driving of the auxiliary system results in an increase in the total mean entropy production rate by an amount $\dot{S}_{\rm exc}$, which captures the thermodynamic cost of operating this information engine.}
    \label{fig:schematic_protocol}
\end{figure}

Here, we {develop such a treatment} by studying a minimal Szilard engine operating autonomously on a single run-and-tumble (RnT) particle in one dimension, {as schematised in Fig.~\ref{fig:schematic_protocol}. We ensure thermodynamic consistency} by expanding the phase space to explicitly include an auxiliary system playing the role of the measurement device. {The auxiliary system's dynamics are designed to induce a correlation between the state of the former and the particle's self-propulsion direction. Positive average work is then readily extracted by applying a time-dependent force smaller than and opposite to the particle’s estimated self-propulsion \cite{pietzonka2019autonomous,di2010bacterial,roberts2023run,cocconi2023optimal}.}
The non-reciprocal coupling between active particle and measurement device introduces the feedback loop which underlies this information engine and the cost of measurement can be computed analytically by evaluating the excess entropy production of the joint dynamics \cite{loos2019non,loos2020irreversibility}. We consider in particular two scenarios: in the first, discussed in Sec.~\ref{s:direct}, the measurement device is a binary process that couples directly to the RnT motility mode; in the second, discussed in Sec.~\ref{s:indirect}, the motility mode is not accessible to direct observation and the measurement device is a continuous state process that couples to the particle position, from whose history the motility mode is inferred. In both cases, the accuracy of the measurement plays a key role in determining the engine's efficiency.

\section{Direct measurement}\label{s:direct}
Consider the one-dimensional run-and-tumble (RnT) process governed by the Langevin equation {for the positional coordinate}
\begin{equation}\label{eq:eom_rnt}
    \dot{x}(t) = \nu w(t) + \gamma^{-1} F_{\rm ext}(t)+ \xi_x(t)
\end{equation}
with $\nu>0$ a self-propulsion speed, $w \in \{-1,1\}$ a symmetric dichotomous noise characterised by a Poisson switching rate $\alpha$, and $\xi_x$ a Gaussian white noise with Dirac-delta covariance, $\langle \xi_x(t)\xi_x(t')\rangle=2 D_x \delta(t-t')$. 
Here, $F_{\rm ext}$ denotes an external force and we work in units such that the viscosity $\gamma=1$. {In practice, $F_{\rm ext}$ could be implemented using an optical trap \cite{sahainfo2023} or, for a charged active colloid \cite{sandoval2016magnetic}, through an external electric field of time-varying magnitude and direction.}
As a preliminary step for the definition of a dynamic Szilard engine operating on a single RnT particle, we expand the phase space to incorporate an auxiliary system, playing the role of a measurement device, which is coupled unidirectionally to the internal state $w$.
We thus introduce a binary Markov process $s \in \{-1,+1\}$ with $w$-dependent transition rate matrix $\Xi_s(w(t))$,
\begin{equation}\label{eq:markov_mat}
    \Xi_s(t) = 
    \begin{cases}
    m
    \begin{pmatrix}
        -\epsilon & 1-\epsilon \\
        \epsilon & -(1-\epsilon)
    \end{pmatrix} &\text{if $w(t)=+1$} \\[5\jot]
    m
    \begin{pmatrix}
        1-\epsilon & \epsilon \\
        -(1-\epsilon) & -\epsilon
    \end{pmatrix} &\text{if $w(t)=-1$}
    \end{cases}
\end{equation}
acting on the vector of probabilities $\mathbf{P}_s(t) = (P_{s=+1}(t),P_{s=-1}(t))$.
The rate $m >0$ in Eq.~\eqref{eq:markov_mat} defines the characteristic timescale of the auxiliary dynamics, while the dimensionless parameter $\epsilon \in [0,1/2]$ can be understood as a measurement error probability. This is a thermodynamically consistent description, in the sense that $ \Delta \mu \equiv \beta^{-1} \ln[(1-\epsilon)/\epsilon]$, with {$\beta^{-1} = k_B T$ a thermal energy scale}, can be interpreted as the free energy drawn from some reservoir to bias the coupling, such that effective decoupling is recovered in the limit $\epsilon \to 1/2$. Indeed, for $\epsilon=1/2$, the auxiliary process $s(t)$ reduces to a symmetric dichotomous noise.

The steady-state joint probability of $w$ and $s$, here denoted $\pi_{w,s}$, can be computed straightforwardly. In particular, we have by symmetry that $\pi_{1,1} = \pi_{-1,-1}$ and $\pi_{1,-1} = \pi_{-1,1}$ with $\pi_{1,1} + \pi_{1,-1} = 1/2$, whence
\begin{align} 
    \pi_{1,1} = \pi_{-1,-1} &= \frac{\alpha + m(1-\epsilon)}{2(2\alpha + m)} \nonumber \\
    \pi_{1,-1} = \pi_{-1,1} &=\frac{\alpha + m\epsilon}{2(2\alpha + m)}    \label{eq:steady_ws_p}
\end{align}
which reduces to $\pi_{\alpha,\beta}=1/4$ for all $\alpha,\beta \in \{1,-1\}$ in the decoupled limit, $\epsilon=1/2$, as expected. 

We will assume the existence of an external controller tasked with the application of the time-dependent external force $F_{\rm ext}(s(t))$, henceforth referred to as the \emph{protocol}, which can be a function of the current state of the measurement device $s(t)$, but not of the hidden self-propulsion state $w(t)$.

\subsection{Excess entropy production}\label{ss:epr_direct}

In order to quantify the efficiency of this dynamic Szilard engine, we define the operational cost of measurement as the excess entropy production rate \cite{seifert2012stochastic,cocconi2020entropy} induced by coupling of the auxiliary degree of freedom $s$ to the RnT dynamics. In isolation, free RnT motion in one dimension is characterised by an entropy production rate $\dot{\mathcal{S}}_{\rm RnT} = \nu^2/D_x$ \cite{cocconi2020entropy}. Upon coupling of the auxiliary system, we write the entropy production rate as the
Kullback-Leibler divergence per unit time of the ensemble
of forward $(x,w,s)$ trajectories of duration $T$ and their time-reversed counterparts to obtain
\begin{align}
    \dot{\mathcal{S}} &\equiv \lim_{T\to\infty} \frac{1}{T}\left\langle \ln \frac{\mathbb{P}_F[x,w,s]}{\mathbb{P}_R[x,w,s]} \right\rangle \nonumber \\
    &= \lim_{T\to\infty} \frac{1}{T} \left[ \left\langle\ln \frac{\mathbb{P}_F[w,s]}{\mathbb{P}_R[w,s]}\right\rangle +\left\langle \ln \frac{\mathbb{P}_F[x|w,s]}{\mathbb{P}_R[x|w,s]} \right\rangle\right]~, \label{eq:epr_direct_full}
\end{align}
with $\mathbb{P}_F$ and $\mathbb{P}_R$ denoting the corresponding path probability densities. The first term in Eq.~\eqref{eq:epr_direct_full} corresponds to the average entropy production rate of the four-state Markov process for the joint dynamics of $w$ and $s$. Using Eqs.~\eqref{eq:markov_mat} and \eqref{eq:steady_ws_p}, this is given by
\begin{align}
    &\lim_{T\to\infty} \frac{1}{T}\left\langle\ln \frac{\mathbb{P}_F[w,s]}{\mathbb{P}_R[w,s]}\right\rangle \nonumber \\
    &= 2[m\epsilon \pi_{1,1} - m(1-\epsilon)\pi_{1,-1}] \ln \frac{m\epsilon \pi_{1,1}}{m(1-\epsilon)\pi_{1,-1}} \nonumber \\
    &\quad  + 2\alpha(\pi_{1,1} - \pi_{-1,1}) \ln \frac{\pi_{1,1}}{\pi_{-1,1}} \nonumber \\
    &= \frac{\alpha m(1-2\epsilon)}{2\alpha + m} \ln \left( \frac{1-\epsilon}{\epsilon} \right)~. \label{eq:epr_ws_direct}
\end{align}
It vanishes at $\epsilon=1/2$, where $w$ and $s$ are independent equilibrium processes, and diverges logarithmically as $\epsilon \to 0$, where state transitions of $s$ become irreversible. 
The second term in Eq.~\eqref{eq:epr_direct_full} can be computed within the Onsager-Machlup path integral formalism \cite{cocconi2020entropy,onsager1953fluctuations}
\begin{small}
\begin{align}
    \mathbb{P}_F[x|w,s] &\propto {\rm exp}\left[ -\frac{1}{4D_x} \int_0^T dt \ \Big( \dot{x}-\nu w(t) - F_{\rm ext}(s(t))\Big)^2 \right] \nonumber \\
    \mathbb{P}_R[x|w,s] &\propto {\rm exp}\left[ -\frac{1}{4D_x} \int_0^T dt \ \left( \dot{x}+\nu w(t) + F^\dagger_{\rm ext}(s(t))\right)^2 \right]
\end{align}
\end{small}
where $F_{\rm ext}^\dagger$ denotes the reversed protocol. Assuming we are working on a ring, such that the marginal probability density $P(x)$ is uniform at steady state, and using the dual-reversed convention \cite{chernyak2006path,seifert2012stochastic} for the time-reversed protocol, $F_{\rm ext}^\dagger(s) = - F_{\rm ext}(s)$, this term reduces to the entropy production $\dot{\mathcal{S}}_{\rm RnT}$ of a free RnT particle. In this instance, the excess entropy $\dot{\mathcal{S}}_{\rm exc} \equiv \dot{\mathcal{S}} - \dot{\mathcal{S}}_{\rm RnT}$ is thus simply given by Eq.~\eqref{eq:epr_ws_direct}. The same result can be obtained by computing the entropy production associated with $x$ in the reference frame {co-transported by the protocol, $\dot{x}_c \equiv \dot{x} - F_{\rm ext}$}. 


\subsection{Na{\"i}ve protocol} \label{ss:naive_direct}
First, let us consider a na{\"i}ve form of the protocol, where the controller applies an external force \emph{as if} the value of the auxiliary system $s$ were an exact copy of the hidden state $w$. It is straightforward to check that, when $w$ is directly observable, the average power output $\langle \dot{W}[F_{\rm ext}]\rangle \equiv -\langle F_{\rm ext}(t) \dot{x}(t)\rangle$ is maximal for $F^*_{\rm ext}(w(t)) = -w\nu/2$, in which case $\dot{W}_{\rm max} \equiv \langle \dot{W}[F^*_{\rm ext}]\rangle=\nu^2/4$ \cite{cocconi2023optimal}. The na{\"i}ve protocol is thus defined by $F_{\rm ext}^{\rm (n)}(s(t))= - {s \nu}/{2}$, resulting in the average power output
\begin{equation}\label{eq:direct_naive_w}
    \langle \dot{W}[F_{\rm ext}^{\rm (n)}]\rangle = \frac{\nu^2}{4}\left[ \langle w s\rangle - \frac{1}{2}\langle s^2 \rangle\right] = \dot{W}_{\rm max} \left[ \frac{m\left( \frac{1}{2} - \epsilon\right) - \alpha}{2\alpha + m} \right]~,
\end{equation}
where we have used
\begin{align} \label{eq:ws_corr_dir}
    \langle w s\rangle &= \pi_{1,1} + \pi_{-1,-1} - \pi_{1,-1} - \pi_{-1,1}  = \frac{m(1-2\epsilon)}{2\alpha + m}
\end{align}
and $s^2(t)= 1$. 
The average power output \eqref{eq:direct_naive_w} is a monotonically decreasing function of $\epsilon$ and, for any finite switching rate $\alpha$, there is a critical error probability $\epsilon_c = (m-2\alpha)/2m$ above which the power output \eqref{eq:direct_naive_w} becomes negative. Furthermore, even in the ideal case $\epsilon=0$, no positive power can be extracted by the application of the na{\"i}ve protocol when $m < 2\alpha$. {The requirement for positive power extraction is more stringent than demanding that the rate of measurement exceeds that of tumbling, reflecting the asymmetry between positive and negative work extracted upon correct and incorrect guessing, respectively.}


Combining the results of Sec.~\ref{ss:epr_direct} with Eq.~\eqref{eq:direct_naive_w} for the average power output, we find the following expression for the {``informational''} efficiency of the conversion from entropy to work of the na{\"i}ve protocol,
\begin{align}
    \eta^{\rm (n)} \equiv \frac{\langle \dot{W}[F_{\rm ext}^{\rm (n)}]\rangle}{\beta^{-1}\dot{\mathcal{S}}_{\rm exc}} &=  \beta \dot{W}_{\rm max} \frac{m\left( \frac{1}{2} - \epsilon\right) - \alpha}{\alpha m(1-2\epsilon) \log \frac{1-\epsilon}{\epsilon}} \nonumber \\
    &= \frac{\beta\dot{W}_{\rm max}}{\log \frac{1-\epsilon}{\epsilon}} \left[ \frac{1}{2\alpha} - \frac{1}{m(1-2\epsilon)}\right] \label{eq:eff_nai_dir}
\end{align}
where $\beta$ is introduced to make the efficiency dimensionless.
Note that the definition of the efficiency $\eta$ used in Eq.~\eqref{eq:eff_nai_dir} and henceforth differs from that of Ref.~\cite{malgaretti2022}, which we denote $\tilde{\eta}$, in that the dissipation of the free active particle is not included in the operational cost. However, they can be related straightforwardly via $\tilde{\eta}(1 + \eta^{-1} + \beta^{-1}\dot{S}_{\rm RnT}/\langle \dot{W}\rangle) = 1$.
We observe that $\eta^{\rm (n)}$ is maximal at an intermediate value of $\epsilon$, \ie at finite power, see Fig.~\ref{fig:power_eff_naive}, and becomes negative for $\epsilon > \epsilon_c$. Since the self-propulsion speed $\nu$ only features via $\dot{W}_{\rm max} \sim \nu^2$, the efficiency can be made arbitrarily large at fixed temperature (constant $\beta$) by increasing $\nu$, in a clear breach of Landauer's bound for equilibrium information engines.  

\subsection{Bayesian protocol}\label{ss:direct_bayesian}

Within the context of direct coupling of the auxiliary process $s$ to the self-propulsion state $w$, let us now consider a more accurate protocol,  which we dub the Bayesian protocol. It was shown in Ref.~\cite{cocconi2023optimal} that maximal average power output from an active particle with hidden states is achieved for a protocol proportional to the posterior expectation of the self-propulsion velocity given the observable degrees of freedom. Denoting $\pi_{w|s} = \pi_{w,s}/\pi_{s}$ the steady-state conditional probabilities, which can be computed from Eq.~\eqref{eq:steady_ws_p}, the Bayesian protocol is accordingly defined by
\begin{equation}\label{eq:opt_force_direct}
    F^{\rm (b)}_{\rm ext}(s(t)) = -\frac{\nu}{2} (\pi_{1|s} - \pi_{-1|s}) = - \frac{s \nu}{2} \frac{m(1-2\epsilon)}{2\alpha + m}~.
\end{equation}
Unlike the na{\"i}ve protocol, Eq.~\eqref{eq:opt_force_direct} depends explicitly on the error probability $\epsilon$, vanishing at $\epsilon = 1/2$ and saturating to $\pm \nu/2$ in the simultaneous limit $\epsilon\to 0$ , $m \to \infty$, at fixed $\alpha$. The associated average power output is
\begin{align}
    &\langle \dot{W}[F^{\rm (b)}_{\rm ext}]\rangle  \nonumber \\
    &= \frac{\nu^2}{2} \left[ \frac{m(1-2\epsilon)}{2\alpha + m} \langle w s\rangle - \frac{m^2 (1-2\epsilon)^2}{2(2\alpha+m)^2} \langle s^2 \rangle  \right] \nonumber \\
    &=  \dot{W}_{\rm max} \frac{m^2(1-2\epsilon)^2}{(2\alpha + m)^2}~, \label{eq:w_bayes_direct}
\end{align}
where we used Eq.~\eqref{eq:ws_corr_dir} to substitute for $\langle w s\rangle$.
Once again, $\dot{W}_{\rm max}$ denotes the power that could have been extracted were $w$ directly accessible to the controller, such that the fraction in the right-hand side of Eq.~\eqref{eq:w_bayes_direct} corresponds to the reduction in performance due to the presence of the measurement device as an intermediary.

\begin{figure}
    \centering
    \includegraphics[width=\columnwidth]{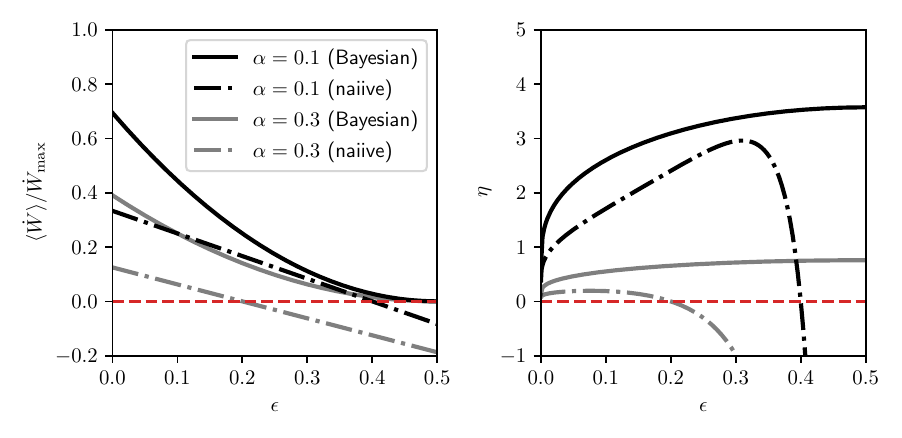}
    \caption{Average power output as a fraction of maximum achievable power (left) and efficiency (right) as a function of the error probability $\epsilon \in [0,1/2]$ for protocols relying on direct coupling to the RnT motility state $w$. Dotted and solid lines refer to results for the na{\"i}ve and Bayesian protocols, as introduced in Sec.~\ref{ss:naive_direct} and \ref{ss:direct_bayesian}, respectively. The power output is a monotonically decreasing function of $\epsilon$ in both cases, with the Bayesian protocol generically outperforming the na{\"i}ve one. The efficiency, on the other hand, {displays a maximum} at finite power for the na{\"i}ve protocol and at zero power for the Bayesian one.
    Here, we set $\dot{W}_{\rm max}=1$ and $m=1$.}
    \label{fig:power_eff_naive}
\end{figure}

Using Eq.~\eqref{eq:epr_ws_direct} for the excess entropy production, the  efficiency of the conversion from entropy to work of the Bayesian protocol is thus
\begin{equation}\label{eq:eff_bayes_indirect}
    \eta^{\rm (b)} \equiv \frac{\langle \dot{W}[F^{\rm (b)}_{\rm ext}]\rangle}{\beta^{-1}\dot{\mathcal{S}}_{\rm exc}} = \frac{\beta \dot{W}_{\rm max} m(1-2\epsilon)}{\alpha(m+4\alpha)\log \frac{1-\epsilon}{\epsilon}}
\end{equation}
which is a monotonically increasing function of $\epsilon$, meaning that in this instance efficiency is maximal in the limit $\epsilon \to 1/2$ and thus at zero power, see Fig.~\ref{fig:power_eff_naive}. In particular, using 
\begin{equation}
    \lim_{\epsilon\to\frac{1}{2}} (1-2\epsilon)\left[ \log \left( \frac{1-\epsilon}{\epsilon} \right) \right]^{-1} = \frac{1}{2}
\end{equation}
we have the value of the max efficiency $\eta^{\rm (b)}_{\rm max}/\beta\dot{W}_{\rm max}  = m/[2\alpha (m+4\alpha)]$. Both work and efficiency vanish in the limit $\alpha \to \infty$, consistent with the idea that the auxiliary system can't ``keep up'' with the $w$ dynamics when the latter's correlation time scale becomes too small. Similarly to the na{\"i}ve protocol studied in the previous Section, the efficiency \eqref{eq:eff_bayes_indirect} can be made arbitrarily large at fixed temperature by increasing $\nu$.\\

\section{Indirect measurement}\label{s:indirect}
In many realistic scenarios, establishing a direct coupling between a measurement device and the internal degree of freedom controlling the self-propulsion direction is unfeasible. For instance, while the position of a molecular motor might be easily tracked, neither the chemical states of its motor heads nor the polarisation of the cytoscheletal filament to which the motor is bound are typically accessible to direct observation \cite{liepelt2007kinesin,neri2022estimating}. In light of this consideration, we now consider a scenario where only the history of the position of the RnT particle is observable. Remarkably, it was shown in Ref.~\cite{cocconi2023optimal} that it is still possible to design an auxiliary system dynamic, such that unidirectional coupling of the latter to the instantaneous particle velocity allows for positive power extraction, the intermediate step being the inference of the hidden state. 

Let us introduce the notation $\{x(t)\}_{-\infty}^\tau$ for the positional trajectory of a RnT particle ending at time $t=\tau$, as well as an auxiliary continuous process $Q[\{x(t)\}_{-\infty}^\tau]$ playing the role of a measurement device, which couples unidirectionally to the {co-transported} particle position. We take, in particular,
\begin{align}
    \dot{Q}(t) 
    &= \frac{\nu}{D_x}[\dot{x}(t)-F_{\rm ext}(t)] - 2\alpha Q + \tilde{\xi}_Q(t) \nonumber \\
    &= \frac{\nu^2}{D_x} w(t) - 2\alpha Q(t) + \xi_Q(t) \label{eq:eom_q}
\end{align}
where $\tilde{\xi}_Q(t)$ and $\xi_Q(t)$ are Gaussian white noises with covariance $\langle \tilde{\xi}_Q(t) \tilde{\xi}_Q(t') \rangle = 2D_Q\delta(t-t')$ and $\langle \xi_Q(t)\xi_Q(t')\rangle = 2(D_Q+\nu^2/D_x) \delta(t-t')$, while $\dot{x}(t)$ was replaced in accordance with Eq.~\eqref{eq:eom_rnt}. 
Thus, $Q(t)$ amounts to a weighted time average of past velocities, with a memory kernel that decays exponentially over a characteristic time scale $1/(2\alpha)$, capturing the intuition that RnT velocities tend to be persistent and of the same sign as $w$ over this time scale.
The non-zero noise correlation $\langle \xi_Q(t)\xi_x(t') \rangle = 2\nu \delta(t-t')$ means that this cannot be thought of as a bipartite system \cite{tanogami2023universal}.
In fact, the limit $D_Q \rightarrow 0$ of Eq.~\eqref{eq:eom_q} can be obtained as the exact low-P{\'e}clet number asymptote [here ${\rm Pe}\equiv \nu^2/(\alpha D_x)$] of the dynamics of the confidence parameter 
\begin{equation}\label{eq:confidence_def}
    \mathfrak{Q}[\{x\}_{-\infty}^\tau] = \ln \frac{P(w(\tau) = +1|\{x\}_{-\infty}^\tau)}{P(w(T) = -1|\{x\}_{-\infty}^\tau)}~,
\end{equation}
where $P(v|\{x\}_{-\infty}^\tau)$ denotes the posterior probability that the internal self-propulsion state is currently $v \in \{1,-1\}$ given the observed positional trajectory \cite{cocconi2023optimal}. Rearranging Eq.~\eqref{eq:confidence_def} as 
\begin{equation}\label{eq:post_prob_from_q}
    P(w(\tau)=\pm 1|\{x\}_{-\infty}^\tau) = \frac{1}{2} \pm \frac{e^\mathfrak{Q}-1}{2(1+e^\mathfrak{Q})}
\end{equation}
clarifies that there exists a one-to-one correspondence between these complementary probabilities and the confidence parameter $\mathfrak{Q}$.
The confidence parameter in turn defines the optimal protocol via the relation
\begin{equation}\label{eq:opt_force_general}
    F_{\rm ext}^*(\mathfrak{Q}(t)) = -\frac{\nu \tanh(\mathfrak{Q}/2)}{2} = -\frac{\nu \mathfrak{Q}}{4} + \mathcal{O}(\mathfrak{Q}^2) ~,
\end{equation}
which draws on the result, already mentioned in Sec.~\ref{s:direct}, that maximal average power output from an active particle with hidden states is achieved for a protocol proportional to the posterior expectation of the self-propulsion velocity given the observable degrees of freedom, which is in this case $x(t)$ \cite{cocconi2023optimal}.

While remaining a sensible intermediary observable to infer $w$, Eq.~\eqref{eq:eom_q} for $Q$ stops being an exact description of the dynamics of the real confidence parameter $\mathfrak{Q}$ for ${\rm Pe} \gtrapprox 1$ and in general when $D_Q >0$. In the following, we will nonetheless define the auxiliary dynamics via Eq.~\eqref{eq:eom_q} at all ${\rm Pe}$, on the basis that no obvious alternative is available. Similarly, the need for a finite measurement noise $D_Q$ in order to establish a well-defined thermodynamic picture of the joint dynamics will become apparent shortly.

\subsection{Excess entropy production}

Following Sec.~\ref{ss:epr_direct} and in order to precisely quantify the efficiency of this dynamic Szilard engine, we define the operational cost of measurement as the excess entropy production rate induced by coupling the auxiliary degree of freedom $Q$ to the RnT dynamics,
\begin{align}
    \dot{\mathcal{S}} 
    &\equiv \lim_{T\to\infty} \frac{1}{T} \left\langle \ln \frac{\mathbb{P}_F[x,w,Q]}{\mathbb{P}_R[x,w,Q]} \right\rangle \nonumber \\
    &= \lim_{T\to\infty} \frac{1}{T} \left\langle \ln \frac{\mathbb{P}_F[x,Q|w]}{\mathbb{P}_R[x,Q|w]} \right\rangle~,
\end{align}
where we have used the fact that a symmetric dichotomous noise $w$ is time-reversal symmetric. 
We write the Onsager-Machlup path probability functional \cite{onsager1953fluctuations} for the forward trajectories in the Stratonovich convention \cite{de2023path,seifert2012stochastic} as
\begin{equation}
    \mathbb{P}_F[x,Q|w] \propto {\rm exp}\left[ - \frac{1}{2} \int_0^T dt \ \mathbf{U}^T(t) \mathsf{C}^{-1} \mathbf{U}(t) \right]
\end{equation}
with
\begin{equation}
    \mathbf{U}(t) = \begin{pmatrix}
\dot{x} - F_{\rm ext}(Q) - \nu w(t) \\
\dot{Q} - \frac{\nu^2}{D_x} w(t) + 2\alpha Q(t)
\end{pmatrix}
\end{equation}
and
\begin{equation}
\mathsf{C}^{-1} = \frac{1}{2 D_x D_Q} 
\begin{pmatrix}
D_Q + \frac{\nu^2}{D_x} & -\nu \\
-\nu & D_x 
\end{pmatrix}
\end{equation}
the inverse covariance matrix.
Similarly, we have for the time-reversed path probability 
\begin{equation}
    \mathbb{P}_R[x,Q|w] \propto {\rm exp}\left[ - \frac{1}{2} \int_0^T dt \ \mathbf{U}_R^T \mathsf{C}^{-1} \mathbf{U}_R \right]
\end{equation}
with
\begin{equation}
    \mathbf{U}_R = \begin{pmatrix}
\dot{x} + F^\dagger_{\rm ext}(Q) + \nu w(t) \\
\dot{Q} + \frac{\nu^2}{D_x} w(t) - 2\alpha Q(t)
\end{pmatrix}~.
\end{equation}
Using the dual-reversed convention \cite{chernyak2006path,seifert2012stochastic} for the time-reversed protocol, $F_{\rm ext}^\dagger(Q) = - F_{\rm ext}(Q)$, and after some algebra, we find 
\begin{equation}
    \dot{S} = \frac{\nu}{D_x} \langle (\dot{x} - F_{\rm ext}) w \rangle + \frac{2\nu\alpha}{D_xD_Q} \langle Q(\dot{x} - F_{\rm ext}) \rangle - \frac{2 \alpha}{D_Q} \langle \dot{Q}Q \rangle~.
\end{equation}
The first term can be identified as the entropy production rate of the free run-and-tumble particle, $\dot{S}_{\rm RnT}$. The remaining two contributions quantify instead the excess entropy originating purely from the coupling to, and nonequilibrium dynamics of, the auxiliary system. Using the steady-state correlations $\langle wQ \rangle={\rm Pe}/4$ and $\langle Q^2 \rangle = D_Q/(2\alpha) + {\rm Pe}/2 + {\rm Pe}^2/8$, which are derived in Appendix \ref{a:tucci} based on Ref.~\cite{Tucci2023},
we finally arrive at the simple expression for the excess entropy production 
\begin{equation}\label{eq:EPR_Bayesian_protocol}
    \dot{S}_{\rm exc} = \frac{\alpha^2{\rm Pe}}{D_Q} \left( 2 + \frac{\rm Pe}{2} \right)+2\alpha~.
\end{equation}
Note that the excess entropy diverges as $D_Q \to 0$, similarly to the limit $\epsilon \to 0$ of the direct measurement setup, Eq.~\eqref{eq:epr_ws_direct}, indicating a vanishing efficiency at finite temperature for the optimal protocol studied in Ref.~\cite{cocconi2023optimal}. Unexpectedly, Eq.~\eqref{eq:EPR_Bayesian_protocol} does not vanish in the limit $D_Q \to \infty$ of infinite measurement noise. We rationalise this result by observing that the correlator $\langle w Q\rangle$ remains finite in this limit, indicating the persistence of a nonequilibrium information flow generated by the coupling between $x$ and $Q$. Finally, unlike Eq.~\eqref{eq:epr_ws_direct}, the excess entropy for the case of indirect measurement depends on $\nu$ via the P{\'e}clet number.

\subsection{Na{\"i}ve protocol}\label{ss:naive_indirect}

\begin{figure}
    \centering
    \includegraphics[width=\columnwidth]{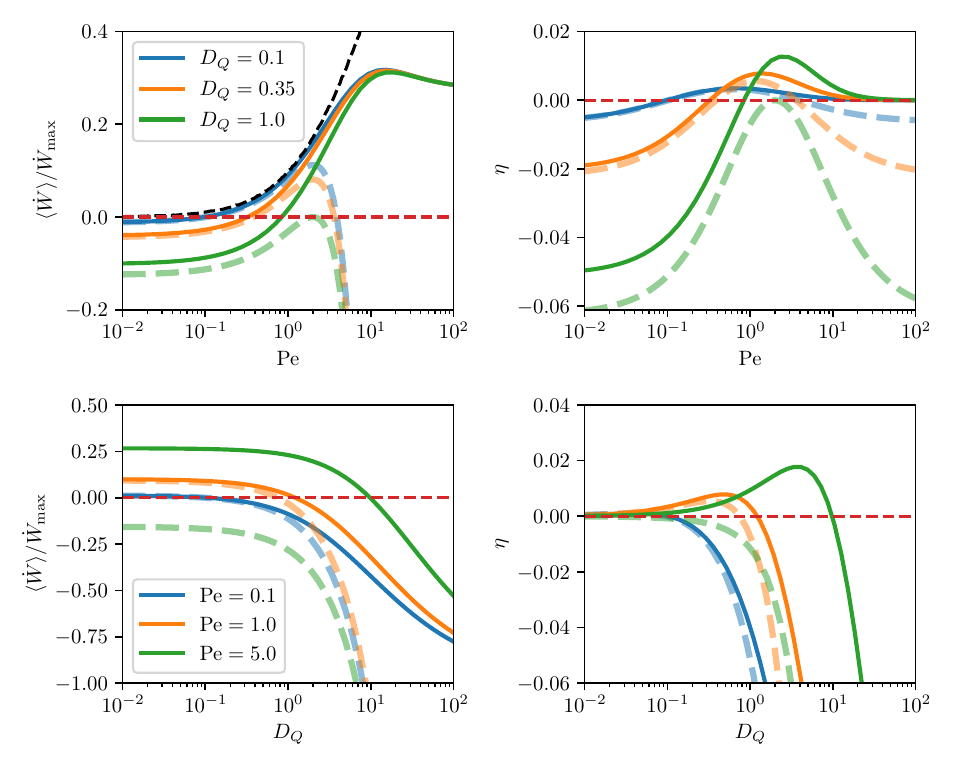}
    \caption{Average power output (left column) and efficiency (right column) as a function of P{\'e}clet number ${\rm Pe} \equiv \nu^2/(\alpha D_x)$ and measurement error $D_Q \in \mathbb{R}^+$ for the na{\"i}ve protocol relying on inference of the RnT motility state $w$ introduced in Sec.~\ref{ss:naive_indirect}, setting $\alpha=\beta=1$. Solid lines denote results for the full nonlinear protocol, Eq.~\eqref{eq:ave_w_ind_naive}  and are obtained by numerical integration. Dashed lines are closed-form analytical results for the linearised protocol, Eqs.~\eqref{eq:ave_w_ind_naive_lowPe} and \eqref{eq:cf_eff_naive}. The linearised protocol performs similarly to the full protocol in the regime $D_Q/\alpha < {\rm Pe} \ll 1$, as expected. In both cases, we observe a non-monotonic dependence of the efficiency on both ${\rm Pe}$ and $D_Q$. The black dashed line in the top-right panel indicates the maximum extractable power under the constraint of a hidden self-propulsion state obtained in Ref.~\cite{cocconi2023optimal}. \label{fig:naive_indirect_1}}
\end{figure}

As was done in Sec.~\ref{ss:naive_direct} for the case of direct measurement, we start with a na{\"i}ve protocol, meaning in this case one that treats the auxiliary system $Q$ as an exact copy of the confidence parameter $\mathfrak{Q}$ of Eq.~\eqref{eq:confidence_def}, ignoring the measurement noise $D_Q$. Accordingly, the na{\"i}ve protocol is defined via Eq.~\eqref{eq:opt_force_general} as $F^{(n)}_{\rm ext}(Q(t)) = F^*_{\rm ext}(\mathfrak{Q}=Q(t))$. 
The associated average power output is
\begin{align}\label{eq:ave_w_ind_naive}
    \langle \dot{W}[F_{\rm ext}^{(n)}] \rangle 
    &= \left\langle \dot{x}(t) \frac{\nu \tanh[Q(t)/2]}{2} \right\rangle~.
\end{align}
The expectation in the right-hand side of Eq.~\eqref{eq:ave_w_ind_naive} can be computed in closed form in the low $\rm Pe$, low $D_Q$ asymptote (specifically $D_Q/\alpha \ll {\rm Pe} \ll 1$). Indeed, we show in Appendix \ref{a:tucci} that $\langle Q^2 \rangle = (D_Q/\alpha + {\rm Pe})/2 + \mathcal{O}({\rm Pe}^2)$, thus in this limit $Q \sim {\rm Pe}$ is typically small and the protocol can be linearised to give
\begin{align}
    \langle \dot{W}[F_{\rm ext,lin}^{(n)}] \rangle &= \left\langle \dot{x} \frac{\nu Q}{4} \right \rangle\nonumber \\
    &= \dot{W}_{\rm max}\left[ \frac{\rm Pe}{8}\left( 1 - \frac{\rm Pe}{4}\right) - \frac{D_Q}{8\alpha}\right]~.
    \label{eq:ave_w_ind_naive_lowPe}
\end{align}
This result, which is exact for all $\rm Pe$ and $D_Q$ but approximates \eqref{eq:ave_w_ind_naive} only when $D_Q/\alpha \ll {\rm Pe} \ll 1$, shows that the fraction of the available power that is actually extracted upon application of the linearised na{\"i}ve protocol is a monotonically decreasing function of the measurement noise $D_Q$, vanishing at a finite critical value $D_{Q,c} = \alpha {\rm Pe}(1-{\rm Pe}/4)$ of the latter. This is also the case for the average power output of the full nonlinear protocol \eqref{eq:ave_w_ind_naive}, as shown in Fig.~\ref{fig:naive_indirect_1} (left column). Furthermore, the power extracted upon application of the linearised protocol eventually becomes negative as $\rm Pe$ is increased. This is not the case for the power output of the full nonlinear protocol \eqref{eq:ave_w_ind_naive}, which instead approaches from above a finite positive value at large $\rm Pe$, see Fig.~\ref{fig:naive_indirect_1} (left column).

\begin{figure}
    \centering
    \includegraphics[width=\columnwidth]{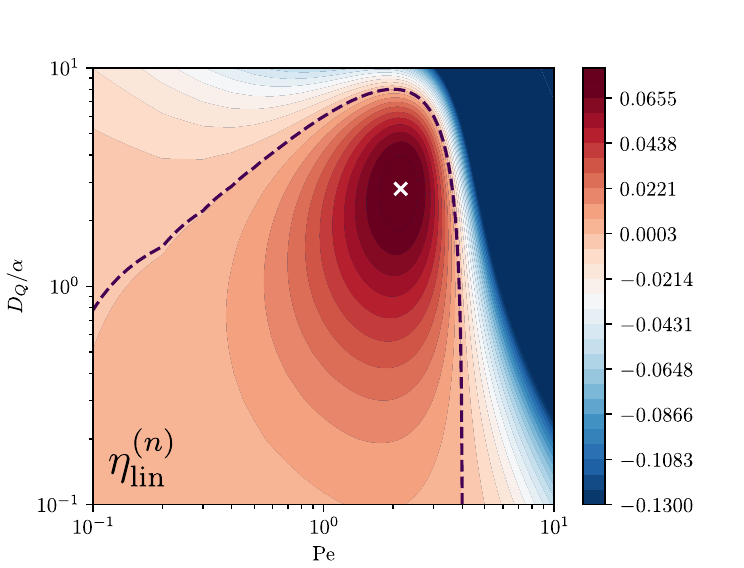}
    \caption{Color map of the efficiency $\eta^{(n)}_{\rm lin}$ of Eq.~\eqref{eq:cf_eff_naive} as a function of the dimensionless parameters $\rm Pe$ and $D_Q/\alpha$ for the linearised na{\"i}ve protocol with indirect measurement discussed in Sec.~\ref{ss:naive_indirect}. The dashed line indicates the contour $\eta^{(b)}_{\rm lin}=0$. The white cross indicates the location of the global maximum, for $\rm Pe = 2.159...$ and $D_Q/\alpha = 2.797...$, where $\eta^{(n)}_{\rm lin} \simeq 0.0794$.}
    \label{fig:contour_eff}
\end{figure}

Combining Eqs.~\eqref{eq:EPR_Bayesian_protocol} and \eqref{eq:ave_w_ind_naive} we finally obtain the efficiency $\eta^{(b)}$ of the conversion from entropy to work of the na{\"i}ve protocol, plotted in Fig.~\ref{fig:naive_indirect_1} (right column). For the linearised protocol, the dimensionless efficiency can be written in closed form,
\begin{equation} \label{eq:cf_eff_naive}
    \eta^{(n)}_{\rm lin} = \frac{
    \beta \dot{W}_{\rm max}\left[ \frac{\rm Pe}{8}\left( 1 - \frac{\rm Pe}{4}\right) - \frac{D_Q}{8\alpha}\right]
    }{2\alpha\left[ \frac{\alpha{\rm Pe}}{D_Q} \left( 1 + \frac{\rm Pe}{4} \right)+1 \right]}~,
\end{equation}
and retains a nonmonotonic dependence on $\rm Pe$ and $D_Q$. Unlike the case of direct measurement studied in Sec.~\ref{s:direct}, here the self-propulsion speed $\nu$ enters both via $\dot{W}_{\rm max} \sim \nu^2$ and ${\rm Pe} \sim \nu^2$ and it is thus not immediately obvious whether $\eta^{(n)}_{\rm lin}$ is bounded above. Assuming that the Stokes-Einstein relation applies, $D_x = k_BT/\gamma$, we can rewrite the efficiency \eqref{eq:cf_eff_naive} as a function of the dimensionless parameters $\rm Pe$ and $D_Q/\alpha$ only. The resulting expression is plotted in Fig.~\ref{fig:contour_eff} and shows a maximum at $\rm Pe = 2.159...$ and $D_Q/\alpha = 2.797...$, where $\eta^{(n)}_{\rm lin} \simeq 8 \times 10^{-2}$, indicating the existence of a nonequilibrium upper bound analogous to that established by Landauer for equilibrium information engines. 

While no closed-form expression is available for the efficiency of the fully nonlinear protocol, it is nevertheless possible to argue that $\eta^{(b)}$ can be made arbitrarily large at fixed temperature. In particular, it is clear that Eq.~\eqref{eq:ave_w_ind_naive} for the average power output is bounded above by $\dot{W}_{\rm max}$ and we further expect $\langle \dot{W}[F_{\rm ext}^{(n)}] \rangle$ to approach a finite fraction of $\dot{W}_{\rm max}$ which is independent of $D_Q/\alpha$ as $\rm Pe \to \infty$, consistently with numerical results in Fig.~\ref{fig:naive_indirect_1}. Thus, for ${\rm Pe} \gg 1$, 
\begin{equation}
    \eta^{(n)} \propto 
    \left[ \frac{\alpha }{D_Q} \left( 1 + \frac{\rm Pe}{4}\right) + {\rm Pe}^{-1} \right]^{-1} ~.
\end{equation}
Taking the limits ${\rm Pe} \to \infty$ and $\alpha {\rm Pe}/D_Q \to 0$ simultaneously produces a divergent efficiency, demonstrating that
$\eta^{(n)}$ is unbounded in $({\rm Pe}, D_Q/\alpha) \in \mathbb{R}^2$. 



\begin{figure}
    \centering
    \includegraphics[width=\columnwidth]{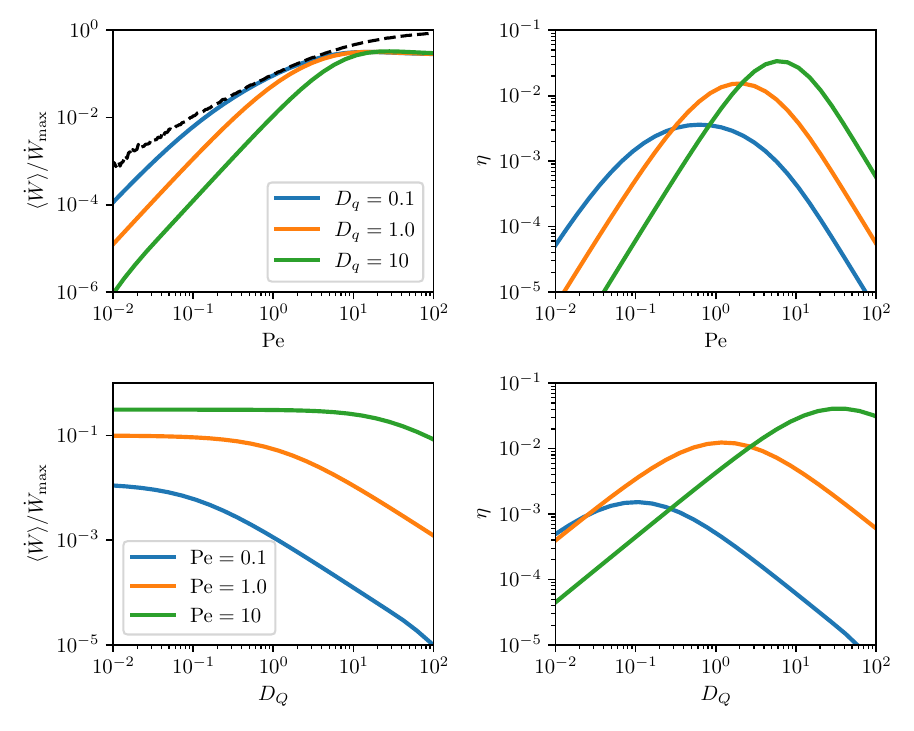}
    \caption{Average power output (left column) and efficiency (right column) as a function of P{\'e}clet number ${\rm Pe} \equiv \nu^2/(\alpha D_x)$ and measurement error $D_Q \in \mathbb{R}^+$ for the Bayesian protocol relying on inference of the RnT motility state $w$ introduced in Sec.~\ref{ss:bayesian_indirect}, Eq.~\eqref{eq:indirect_bayes_prot}. The results are obtained by numerical integration of exact expressions, setting $\alpha=\beta=1$, for which no closed form is available. Similarly to the na{\"i}ve protocol, we observe a non-monotonic dependence of the efficiency on both ${\rm Pe}$ and $D_Q$. The black dashed line in the top-left panel indicates the maximum extractable power under the constraint of a hidden self-propulsion state (data from Ref.~\cite{cocconi2023optimal}).}
    \label{fig:bayes_1}
\end{figure}

\subsection{Bayesian protocol}\label{ss:bayesian_indirect}

Completing the parallel with Sec.~\ref{s:direct}, we now consider a more accurate ``Bayesian'' protocol designed to account and partially compensate for the measurement noise parametrised by $D_Q$. Rather than treating the auxiliary variable $Q$ as a copy of the confidence parameter $\mathfrak{Q}$ of Eq.~\eqref{eq:confidence_def}, which is related to the posterior probability $P(w(\tau)|Q(\tau))$ of the self-propulsion mode via the simple relation \eqref{eq:post_prob_from_q}, we compute the true self-propulsion probabilities conditioned on $Q$ explicitly. In particular, we write 
\begin{align}
    &P(w(\tau)=\pm 1|Q(\tau)) \nonumber \\
    &= \int d\mathfrak{Q} P(w(\tau)=\pm 1|\mathfrak{Q})P(\mathfrak{Q}|Q) \nonumber \\
    &= \int d\mathfrak{Q} \left[ \frac{1}{2} \pm \frac{e^\mathfrak{Q}-1}{2(1+e^\mathfrak{Q})} \right] \frac{P(Q|\mathfrak{Q}) P(\mathfrak{Q})}{P(Q)}
\end{align}
where we have used Bayes' theorem together with Eq.~\eqref{eq:post_prob_from_q}. By inspection of the Langevin equation \eqref{eq:eom_q}, it is clear that the steady state probability $P(Q)$ is identical to that of the position of a RnT particle confined in a harmonic potential of stiffness $2\alpha$, which is known in closed form from Refs.~\cite{Tucci2023,garcia2021run}, see Appendix \ref{a:tucci}. On the other hand, neither $P(Q|\mathfrak{Q})$ nor $P(\mathfrak{Q})$ are known analytically. Nevertheless, we can draw on the result of Ref.~\cite{cocconi2023optimal} that the dynamics of $\mathfrak{Q}$ are well approximated at low $\rm Pe$ by Eq.~\eqref{eq:eom_q} with $D_Q=0$ to write $P(\mathfrak{Q}) \simeq P(\tilde{Q})$ and $P(Q|\mathfrak{Q}) \simeq P(Q|\tilde{Q})$, where
\begin{equation} \label{eq:Q_OU_correction}
    Q(t) = \tilde{Q}(t) + \int_{-\infty}^\tau dt \ \tilde{\xi}_Q(t) e^{-2\alpha(\tau-t)}~.
\end{equation}
The second term in the right-hand side of Eq.~\eqref{eq:Q_OU_correction} is simply an Ornstein-Uhlenbeck process \cite{OUoriginal} with zero mean and variance $D_Q/2\alpha$, implying $P(Q|\tilde{Q}) = \mathcal{N}(Q;\tilde{Q},D_Q/2\alpha)$, where $\mathcal{N}(\bullet;m,v)$ denotes a Gaussian distribution of mean $m$ and variance $v$. Our Bayesian protocol is then defined as
\begin{align}
    &F_{\rm ext}^{\rm (b)}(Q(t)) \nonumber \\
    &= - \frac{\nu}{2} [P(w(t)=+1|Q(t)) - P(w(t)=-1|Q(t))] \nonumber \\
    &= -\frac{\nu}{2} \int d\tilde{Q} \left( \frac{e^{\tilde{Q}}-1}{e^{\tilde{Q}}+1} \right) \mathcal{N}(Q;\tilde{Q},D_Q/2\alpha) \frac{P(\tilde{Q})}{P(Q)}~.
    \label{eq:indirect_bayes_prot}
\end{align}
The average power output associated with this protocol can be computed by numerical evaluation of the exact integral expressions, giving the curves shown in Fig.~\ref{fig:bayes_1} (left column). These results can then be combined with Eq.~\eqref{eq:EPR_Bayesian_protocol} for the excess entropy production to compute the efficiency $\eta^{\rm (b)}$, also shown in Fig.~\ref{fig:bayes_1} (right column). 
{The non-monotonic dependence of $\eta^{\rm (b)}$ on ${\rm Pe}$, which reflects the onset of a regime where measurement expenditure exceeds power output, is reminiscent of recent results in quantum Szilard engine with finite-time measurement \cite{zhou2023finitetime} and partially observable information engines \cite{Still2020memory}.}
While no closed-form expression can realistically be obtained for the average power output associated with the protocol \eqref{eq:indirect_bayes_prot}, a similar logic to the one presented in the previous section can be used to argue for the corresponding efficiency to be unbounded in $({\rm Pe}, D_Q/\alpha) \in \mathbb{R}^2$. Indeed, such an argument relies solely on the assumption, consistent with our numerical results, that the average power output should tend to a finite constant independent of $D_Q/\alpha$ as $\rm Pe \to \infty$.


\section{Conclusion}
We have addressed the thermodynamic consistency of an autonomous, dynamic Szilard engine \cite{malgaretti2022} operating on a RnT particle in one dimension, a canonical model in active matter physics reminiscent of the motility strategy of \emph{E.Coli} and \emph{Salmonella} bacteria \cite{elgeti2015physics,garcia2021run,slowman2017exact,renadheer2019path,solon2015active}. 
We have considered two realistic, \ie error-prone, measurement scenarios: in the first, binary measurements are performed directly on the internal self-propulsion state of the active particle, while in the second only the particle trajectory in space is accessible and the self-propulsion state is estimated by means of inference from weighted time-averages of past velocities \cite{cocconi2023optimal}. 
In each scenario, {we contrast a na{\"i}ve version of the protocol, where measurement errors are neglected, with a Bayesian version, where knowledge of the error statistics is drawn upon to generically enhance efficiency.}
The thermodynamic cost of operating the measurement device is quantified formally as the excess entropy production rate induced by the coupling of an auxiliary process {playing the role of a measurement device}, which we compute analytically and in closed form. 
{We showed that, in both cases, the precision of measurements has a nontrivial impact on the efficiency of the engine.}
With the exception of the linearised na{\"i}ve protocol studied in Sec.~\ref{ss:bayesian_indirect}, we observed that in all instances the efficiency of our dynamic Szilard engine can be made arbitrarily large at constant temperature by increasing the dissipation of the free RnT particle, consistent with recent results \cite{malgaretti2022} where operational costs were estimated in a more heuristic way. We understand this breach of Landauer's equilibrium bound as a consequence of information being used to redirect part of the intrinsic internal dissipation originating from activity into useful work, rather than to extract the latter directly from a heat bath with an infinitesimal correlation time scale.

The analysis presented here could be refined by considering the efficiency of a similarly designed autonomous information engine operating on a thermodynamically consistent active particle model, such as those introduced in Refs.~\cite{liepelt2007,chatzittofi2023entropy,fritz2023thermodynamically,bebon2024thermodynamics}, which typically involve a weak coupling between configurational/chemical and mechanical degrees of freedom, with multiple dissipative currents, rather than describing self-propulsion via an effective external ``active force" \cite{shankar2018,speck2016stochastic}.\\

Both LCs acknowledge Connor Roberts, Yu-Han Ma and Henry Alston for useful feedback on the manuscript. \\

\appendix
\section{Statistics of RnT motion in a harmonic trap}\label{a:tucci}
The authors of Ref.~\cite{Tucci2023} have obtained a simple integral expression for the steady-state probability density of a generic active particle in a harmonic potential, on which we draw in various places of this work for the particular case of RnT motion and thus report here for convenience. In particular, they
considered an Ornstein-Uhlenbeck process with time-dependent potential described by the Langevin equation
\begin{equation}\label{eq:Tucci}
    \dot{x}(t) = -\frac{k}{\gamma}\left [x(t)-c(t)\right ]+\xi(t)~
\end{equation}
where $k$ is the stiffness of the harmonic potential $V(x,c)=k(x-c)^2/2$ centered at $c(t)$, $\gamma$ denotes the effective friction coefficient and $\xi(t)$ is a Gaussian white noise with zero mean and covariance $\langle \xi(t_1) \xi(t_2) \rangle = 2 D\delta(t_1-t_2)$. By the Stokes-Einstein relation, the effective diffusion coefficient satisfies $D=k_B T/\gamma$. RnT motion is captured by Eq.~\eqref{eq:Tucci} when centre $c(t)$ is a symmetric dichotomous process taking the values $\pm c_0$ with switching rate $r$.  With these definitions and setting $\gamma=1$ without loss of generality, the stationary conditional distributions of $x$, given $\sigma=c/c_0 = \pm 1$, reads
\begin{align}
    P(x|\sigma) &= \frac{N}{2}\int_{-1}^{+1} dz\ \mathcal{N}\left(x;c_0 z,\frac{D}{\nu}\right)(1-\sigma z)^{r/\nu-1}(1+\sigma z)^{r/\nu} ~,\label{eq:Tucci_results_rho} 
\end{align}
where $\mathcal{N}(\bullet;m,v)$ once again denotes a Gaussian distribution of mean $m$ and variance $v$ and
\begin{align}
     N^{-1} &= \frac{2\nu}{r} {}_{2}F_{1}\left(1,1-\frac{r}{\nu}, 1+\frac{r}{\nu},-1 \right)\label{eq:Tucci_results_N}
\end{align}
is a normalisation factor. 
Here, $\nu = {k}/{\gamma}$, while ${}_{2}F_{1}$ denotes the hypergeometric function.

For the $Q$ dynamics of Eq.~\eqref{eq:eom_q}, we thus have,
by direct comparison with \eqref{eq:Tucci} and incorporating results of Eqs.~\eqref{eq:Tucci_results_rho} and \eqref{eq:Tucci_results_N}, the steady-state conditional probability density

\begin{align}
    P(Q|w) = \frac{N_Q}{8}\int_{-1}^{+1} dz\ 
    \mathcal{N}&\left( Q; \frac{\nu^2}{2\alpha D_x} z, \frac{1}{2\alpha}\left( \frac{\nu^2}{D_x} + D_Q\right)\right) \nonumber \\
    &\qquad \times (1-w z)^{-\frac{1}{2}}(1+w z)^{\frac{1}{2}} \label{eq:cond_p_qw}
\end{align}
with $N_Q^{-1} = {}_{2}F_{1}(1,1/2, 3/2,-1)$.
Using Eq.~\eqref{eq:cond_p_qw} is is straightforward to derive the expectations 
\begin{equation}
    \langle Q^2 \rangle = \frac{D_Q}{2\alpha} + \frac{\rm Pe}{2}\left( 1 + \frac{\rm Pe}{4}\right),\quad \langle wQ\rangle = \frac{\rm Pe}{4}~.
\end{equation}
The corresponding statistics of $\tilde{Q}$, as introduced in Eq.~\eqref{eq:Q_OU_correction}, are obtained from the above by simply setting $D_Q=0$. Equivalent results have been obtained in Ref.~\cite{garcia2021run}, albeit in a different representation.


\bibliography{sample}

\end{document}